\documentclass[12pt]{article}
\usepackage{graphicx}
\usepackage{amsfonts}
\usepackage{amssymb}

\def\hybrid{\topmargin -20pt    \oddsidemargin 0pt
        \headheight 0pt \headsep 0pt
        \textwidth 6.25in
        \textheight 9.00in       
        \marginparwidth .875in
        \parskip 5pt plus 1pt   \jot = 1.5ex}

\hybrid

\catcode`\@=11

\def\marginnote#1{}
\newcount\hour
\newcount\minute
\newtoks\amorpm
\hour=\time\divide\hour by60
\minute=\time{\multiply\hour by60 \global\advance\minute by-\hour}
\edef\standardtime{{\ifnum\hour<12 \global\amorpm={am}%
        \else\global\amorpm={pm}\advance\hour by-12 \fi
        \ifnum\hour=0 \hour=12 \fi
        \number\hour:\ifnum\minute<10 0\fi\number\minute\the\amorpm}}
\edef\militarytime{\number\hour:\ifnum\minute<10 0\fi\number\minute}

\def\draftlabel#1{{\@bsphack\if@filesw {\let\thepage\relax
   \xdef\@gtempa{\write\@auxout{\string
      \newlabel{#1}{{\@currentlabel}{\thepage}}}}}\@gtempa
   \if@nobreak \ifvmode\nobreak\fi\fi\fi\@esphack}
        \gdef\@eqnlabel{#1}}
\def\@eqnlabel{}
\def\@vacuum{}
\def\draftmarginnote#1{\marginpar{\raggedright\scriptsize\tt#1}}

\def\draft{\oddsidemargin -.5truein
        \def\@oddfoot{\sl preliminary draft \hfil
        \rm\thepage\hfil\sl\today\quad\militarytime}
        \let\@evenfoot\@oddfoot \overfullrule 3pt
        \let\label=\draftlabel
        \let\marginnote=\draftmarginnote
   \def\@eqnnum{(\theequation)\rlap{\kern\marginparsep\tt\@eqnlabel}%
\global\let\@eqnlabel\@vacuum}  }

\def\preprint{\twocolumn\sloppy\flushbottom\parindent 2em
        \leftmargini 2em\leftmarginv .5em\leftmarginvi .5em
        \oddsidemargin -.5in    \evensidemargin -.5in
        \columnsep .4in \footheight 0pt
        \textwidth 10.in        \topmargin  -.4in
        \headheight 12pt \topskip .4in
        \textheight 6.9in \footskip 0pt
        \def\@oddhead{\thepage\hfil\addtocounter{page}{1}\thepage}
        \let\@evenhead\@oddhead \def\@oddfoot{} \def\@evenfoot{} }

\def\numberbysection{\@addtoreset{equation}{section}
        \def\theequation{\thesection.\arabic{equation}}}

\def\underline#1{\relax\ifmmode\@@underline#1\else
        $\@@underline{\hbox{#1}}$\relax\fi}

\def\titlepage{\@restonecolfalse\if@twocolumn\@restonecoltrue\onecolumn
     \else \newpage \fi \thispagestyle{empty}\c@page\z@
        \def\thefootnote{\fnsymbol{footnote}} }

\def\endtitlepage{\if@restonecol\twocolumn \else \newpage \fi
        \def\thefootnote{\arabic{footnote}}
        \setcounter{footnote}{0}}  

\catcode`@=12
\relax

\def\figcap{\section*{Figure Captions\markboth
        {FIGURECAPTIONS}{FIGURECAPTIONS}}\list
        {Figure \arabic{enumi}:\hfill}{\settowidth\labelwidth{Figure
999:}
        \leftmargin\labelwidth
        \advance\leftmargin\labelsep\usecounter{enumi}}}
 \relax
\def\tablecap{\section*{Table Captions\markboth
        {TABLECAPTIONS}{TABLECAPTIONS}}\list
        {Table \arabic{enumi}:\hfill}{\settowidth\labelwidth{Table
999:}
        \leftmargin\labelwidth
        \advance\leftmargin\labelsep\usecounter{enumi}}}
 \relax
\def\reflist{\section*{References\markboth
        {REFLIST}{REFLIST}}\list
        {[\arabic{enumi}]\hfill}{\settowidth\labelwidth{[999]}
        \leftmargin\labelwidth
        \advance\leftmargin\labelsep\usecounter{enumi}}}
 \relax
\makeatletter
\newcounter{pubctr}
\def\publist{\@ifnextchar[{\@publist}{\@@publist}}
\def\@publist[#1]{\list
        {[\arabic{pubctr}]\hfill}{\settowidth\labelwidth{[999]}
        \leftmargin\labelwidth
        \advance\leftmargin\labelsep
        \@nmbrlisttrue\def\@listctr{pubctr}
        \setcounter{pubctr}{#1}\addtocounter{pubctr}{-1}}}
\def\@@publist{\list
        {[\arabic{pubctr}]\hfill}{\settowidth\labelwidth{[999]}
        \leftmargin\labelwidth
        \advance\leftmargin\labelsep
        \@nmbrlisttrue\def\@listctr{pubctr}}}
 \relax
\makeatother
\newskip\humongous \humongous=0pt plus 1000pt minus 1000pt

\newif\ifdtup

\relax

\def\be{\begin{equation}}
\def\ee{\end{equation}}
\def\ba{\begin{eqnarray}}
\def\ea{\end{eqnarray}}

\def\IR{\relax{\rm I\kern-.18em R}}

\def\IR{\relax{\rm I\kern-.18em R}}
\def\inv{^{\raise.15ex\hbox{${\scriptscriptstyle -}$}\kern-.05em 1}}

\begin{document}


\renewcommand{\theequation}{\arabic{equation}}

\newcommand{\beq}{\begin{equation}}
\newcommand{\eeq}[1]{\label{#1}\end{equation}}
\newcommand{\ber}{\begin{eqnarray}}
\newcommand{\eer}[1]{\label{#1}\end{eqnarray}}
\newcommand{\eqn}[1]{(\ref{#1})}


\begin{titlepage}
\begin{center}
\hfill hep--th/0112113\\
\vskip .05in
\hfill CALT-68-2362\\
\vskip .05in
\hfill CITUSC-01-048\\
\vskip .6in
{\Large \bf $G_2$ Holonomy Spaces from\\[12pt] Invariant Three-Forms}
\vskip .4in
{\large Andreas Brandhuber}
\vskip .4in
{\em Department of Physics\\
California Institute of Technology\\
Pasadena, CA 91125, USA\\[12pt]
and\\[12pt]
Caltech-USC Center for Theoretical Physics\\
University of Southern California\\
Los Angeles, CA 90089, USA}\\
\end{center}
\vskip .6in
\centerline{\bf Abstract}
\vskip .2in
\noindent
We construct several new $G_2$ holonomy metrics that play an important
role in recent studies of geometrical transitions in compactifications
of M-theory to four dimensions.
In type IIA string theory these metrics correspond
to D6 branes wrapped on the three-cycle of the deformed conifold
and the resolved conifold with
two-form RR flux on the blown-up two-sphere, which are
related by a conifold transition. We also study a $G_2$ metric that
is related in type IIA to the line bundle over ${\bf S}^2 \times {\bf S}^2$
with RR two-form flux. Our approach exploits systematically
the definition of torsion-free $G_2$ structures in terms of
three-forms which are closed and co-closed. Besides being an elegant
formalism this turns out to be a practical tool to construct $G_2$
holonomy metrics.
\vskip .3in
\noindent
December 2001\\
\end{titlepage}
\vfill
\eject

\def\baselinestretch{1.2}
\baselineskip 16 pt
\noindent


\section{Introduction}

Compactifications of M-theory on seven-manifolds with $G_2$ holonomy
have recently attracted increased attention. If the seven-manifold is smooth
the low energy theory contains only four-dimensional $N=1$ supergravity
coupled to a number of $U(1)$ vector multiplets and neutral chiral
multiplets \cite{paptow}.
Due to the lack of non-abelian gauge symmetries and chiral matter it might
seem that such compactifications are physically uninteresting but
dualities imply that the situation cannot be as dire. Indeed, it has
recently be found that both non-abelian gauge symmetries and chiral fermions
can be included if we allow the seven-manifold to be singular
\cite{ach1,amv,atiwit,witten,achwit}.

Non-abelian gauge symmetries can be explained most easily by fiberwise
application of the duality between heterotic string on ${\bf T}^3$ and
M-theory on $K3$. Enhanced gauge symmetry on the heterotic side
translates to a singular limit of $K3$ on the M-theory side where an
$ADE$ singularity appears corresponding to $ADE$ gauge groups. Fibering
this over a compact three-cycle produces examples of $G_2$ holonomy spaces.
Duals of $N=1$ supersymmetric gauge theories based on orbifolds of
known $G_2$ metrics on the spin bundle of the three-sphere
${\bf R^4}/\Gamma_{ADE} \times {\bf S}^3$ were studied
in \cite{ach1,amv}. Furthermore, it was suggested that after the
singular $G_2$ manifold undergoes a flop transition it is replaced by
the smooth orbifold ${\bf R^4} \times {\bf S}^3/\Gamma_{ADE}$
\cite{ach1,amv}
and describes the strong coupling regime of $N=1$ SYM \cite{ach1,amv}.
In particular, the existence of a mass-gap, $\chi$SB and
confining strings charged under the center of the gauge group
can be identified in the supergravity dual \cite{amv,atiwit,ach2}.
Note that such singularities appear in codimension four and are not
special to $G_2$ spaces, they also appear in Calabi-Yau three- and
four-folds. These types of dualities and phase transitions have been
further studied and generalized in \cite{ach2} - \cite{Bilal}.

More exotic are the codimension seven (pointlike) singularities that
give rise to chiral matter \cite{witten}.
On one hand they can be understood via
duality with type IIA string theory where the singularity
corresponds to the point where stacks of parallel D6 branes
intersect \cite{atiwit}.
Alternatively they can be described in heterotic string
theory on Calabi-Yau three-folds in which the rank of the gauge bundle
jumps over isolated points on the three-fold \cite{achwit}.

In this paper we provide a general formalism to construct
$G_2$ holonomy metrics which are of interest to improve
our understanding of
the above mentioned dualities and geometric transitions.
The approach we use differs from the conventional procedures
which usually starts from an ansatz for the metric.
Here we exploit the mathematical fact that a torsion-free
$G_2$ structure is characterized by an invariant three-form which is
closed and co-closed and the corresponding $G_2$ metric
is a non-linear function of the three-form
\cite{bryant,bs,joyce,Cleyton}.
(See also \cite{hitchin} for a related approach.)
Hence, our starting point will be an ansatz for
the three-form which incorporates all symmetries we wish to
impose. This method has the advantage that it reduces the tangent
space symmetry $GL(7)$ directly down to $G_2$, and not in
a two step process first to $SO(7)$ for the metric ansatz
and then down to $G_2 \subset SO(7)$ by imposing $G_2$. When using
the latter approach solutions can be missed if the metric
ansatz is incompatible with the $G_2$ structure one wants to
impose.
The formalism we present is not only of academic interest,
since it turns out to be a practical tool to construct
$G_2$ holonomy metrics.
The ansatz for the three-form requires only a minimal number of
unknown functions and it is straightforward, though in general
very tedious, to determine the condition for $G_2$ holonomy.
In general this condition boils down to a set of highly
non-linear differential equations which can be solved exactly
only in special cases and have to be studied numerically otherwise.
New examples of metrics with $G_2$ holonomy have been constructed
recently \cite{bggg} \cite{cglp} \cite{cvnew} \cite{Cvetic:2001zx}
\cite{achwit}. Also the search for $Spin(7)$ metrics was revived
recently \cite{cglp1}-\cite{Kanno:2001xh}.

The plan of the paper is as follows.
In section 2 we show how the fact that $G_2$ holonomy metrics
are characterized by a three-form that is closed and co-closed
can be used to construct metrics with torsion-free $G_2$ structure.
We illustrate this on a known example where the metric is
asymptotic to a cone over ${\bf S}^3 \times \tilde{\bf S}^3$
by writing down the most general three-form ansatz compatible
with $SU(2)^3$ symmetry. In this way we find the most general $G_2$
metric with this symmetry and by requiring in addition regularity
we reproduce the three previously known asymptotically conical (AC) metrics.
In section 3 we systematically search for new metrics by
constructing the most general three-form ansatz that
preserves an $SU(2)^2 U(1)$ symmetry.
Using our method we find three metrics that correspond to
Kaluza-Klein monopoles fibered over one of three possible
three-spheres, one of which was constructed recently
\cite{bggg}. In type IIA string theory these backgrounds correspond
to a configuration of one (several) D6-brane(s) wrapped on the
three-sphere inside a deformed conifold.
Furthermore, we find a completely new class of
metrics that have the interpretation of being
the M-theory lift of the small resolution of the conifold with
one (or more) unit(s) of RR two-form flux on the blown up two-sphere.
This metric is
interesting because it is related in weakly coupled
string theory to the wrapped
D6 brane solution via Vafa's conifold transition \cite{vafa}
and is the supergravity dual of $N=1$ SYM at strong coupling
\cite{ach1,amv,ach2}. The M-theory lift
at infinite string coupling, on the other hand, was described
in \cite{amv} using the $SU(2)^3$ symmetric $G_2$ spaces
\cite{bs,gpp}. Hence,
the solutions we find describe the interpolation between these two
pictures for arbitrary, finite string coupling.
Furthermore, this new metric has the novel feature that
it has an $U(1)$ isometry whose orbit is finite everywhere. It
never blows up or shrinks to zero so the metric in M-theory and
after reduction to type IIA the ten-dimensional solution is
non-singular everywhere even when the number of units of RR flux
is larger than one. Asymptotically, the metric is similar to
the {\cal brane} solution and corresponds to a $U(1)$ fibration
over the conifold with a finite size fiber.
A metric with similar features but a different blown up cycle and
different asymptotics has been found recently \cite{cvnew}.
It is a new branch of solutions of the equations that were first
found in \cite{bggg} in the search for new $G_2$ metrics.
We present further numerical evidence for the existence of this
solution. Finally, in section 4 we conclude with
a discussion of our results and remarks on generalizations
of our formalism. In Appendix A we present the general
$SU(2)^2 U(1)$ symmetric three-form ansatz and the corresponding
metric and condition for $G_2$ holonomy.
Appendix B summarizes similar information for a related ansatz
with $SU(2)^2$ symmetry.

\section{$G_2$ Holonomy Manifolds}

$G_2$ holonomy metrics on a seven-manifold $X$ are solutions of
eleven-dimensional supergravity and can be used to describe
four-dimensional vacua with $N=1$, $d=4$ supersymmetry of the
type ${\bf R}^{1,3} \times X$. If $X$ has finite volume the
four-dimensional Newton constant is finite, but since we will
exclusively study non-compact $X$ gravity lives in eleven
dimensions.
Another important ingredient of the metrics we study is that they have
at least an $U(1)$ isometry\footnote{Note that compact
examples in general lack any continuous symmetries.}
which allows to reduce the purely geometric background
${\bf R}^{1,3} \times X$ in
$M$-theory to a background in type IIA string theory using
\be\label{m2a}
ds_{11}^2 = e^{-2 \phi/3} ds_{10}^2 + e^{4 \phi/3}
\Big( dx_{11} + C_\mu dx^\mu \Big)^2
\ee
where $\phi$ and $C$ are the type IIA dilaton and Ramond-Ramond (RR)
one-form, respectively. In type IIA string theory these backgrounds
involve intersecting D6 branes, D6 branes wrapped on supersymmetric
cycles of Calabi-Yau three-folds, or RR two-form fluxes $F=dC$ over
non-trivial cycles of three-folds. For numerous examples of
these types, see \cite{amv,atiwit,achwit,bggg,cglp1,cvnew}.

We want to find an effective method to construct new metrics with
$G_2$ holonomy on seven-manifolds $X$.
As mentioned above we focus on non-compact examples, which are
important to study $M$-theory on compact $G_2$ manifolds in the
vicinity of singularities. Physical transitions can occur if the
singularities are resolved in inequivalent ways.
In many cases these compactifications also provide interesting
supergravity duals of supersymmetric gauge theories
\cite{ach1,ach2,atiwit} similar to \cite{ks,manu}.

In the context of compactifications of $M$-theory the condition of
$G_2$ holonomy is simply the condition of $d=4$, ${\cal N}=1$
supersymmetry and requires the existence of precisely one
covariantly constant spinor. For practical purposes this condition
is not very helpful and we will use different, mathematically
equivalent conditions. For this purpose let us first review some
basic mathematical facts about $G_2$ structures and metrics of
$G_2$ holonomy.
In flat ${\bf R}^7$ with metric
\be\label{flatmeric}
ds^2 = dx_1^2 + \ldots + dx_7^2
\ee
the Lie group $G_2$, which is also
the invariance group of the unit octonions, can be defined as the
stabilizer of the three-form
\be\label{3form}
\Phi_0 = dx_{123} + dx_{147} + dx_{165} + dx_{246} +
dx_{257} + dx_{354} + dx_{367}
\ee
and the four-form
\be\label{4form}
\ast \Phi_0 = dx_{4567} + dx_{2356} + dx_{2374} + dx_{1357} +
dx_{1346} + dx_{1276} + dx_{1245}
\ee
under the natural action of $GL(7,{\bf R})$.
In other words $\Phi_0$ and
$\ast \Phi_0$ define a $G_2$ structure on ${\bf R}^7$.
On a general curved manifold $X$ a $G_2$ structure is an identification
of the tangent space ${\cal T}X$ with the unit quaternions.
Equivalently, the geometry is determined by a {\it stable} three-form
$\Phi$ for which at every point $p \in X$ there exists an
isomorphism between ${\cal T}_p X$ and ${\bf R}^7$ that identifies
$\Phi$ with $\Phi_0$ in \eqn{3form}.
Stability in the sense of Hitchin \cite{hitchin} means that $\Phi$
lies in a particular open orbit of the action of $GL(7,{\bf R})$.
Most importantly for us, $\Phi$ determines a metric $g_{ij}$ on
$X$ and hence a Hodge-star operator $\ast$.
Furthermore, if $\Phi$ and $\ast \Phi$ are closed, then the metric
$g_{ij}$ is Ricci flat and has holonomy $\mathrm{Hol}(g) \subseteq G_2$.
If in addition $\pi_1(X)$ is finite or $b_1(X)=0$ then
$\mathrm{Hol}(g) = G_2$.

The following two equivalent definitions of $G_2$ holonomy are
interesting for us:
\begin{itemize}
\item
The $G_2$ structure $(\Phi,g)$ is torsion-free:
\be\label{covconst}
\nabla \Phi = 0 ~.
\ee
\item
The three-form $\Phi$ is closed and co-closed:
\be\label{cclosure}
d \Phi = 0 \quad \mathrm{and} \quad d \ast_\Phi \Phi = 0 ~.
\ee
\end{itemize}
We will use the latter definition,
hence, the construction of $G_2$ holonomy metrics on seven-manifolds is
equivalent to the construction of globally defined
three-forms $\Phi$ that are closed and co-closed. Note that in the
compact case this is equivalent to the condition that $\Phi$ be harmonic,
however, in the non-compact case the two are not equivalent and we have
to use the stronger condition \eqn{cclosure}.
The $G_2$ holonomy metric can be expressed in terms of the three-form
\cite{bryant,bs}
\ba\label{g2metric}
g_{ij} & = & \left( \det s_{ij}\right)^{-1/9} s_{ij} ~, \nonumber \\
s_{ij} & = & -\frac{1}{144}\Phi_{i m_1 m_2} \Phi_{j m_3 m_4}
\Phi_{m_5 m_6 m_7}
\epsilon^{m_1 \ldots m_7} ~,~ \epsilon^{1234567} = +1 ~.
\ea
Although \eqn{covconst} and \eqn{cclosure} appear linear in $\Phi$,
in fact $\ast_\Phi$ and $\nabla$ depend on the metric $g$, which depends
on $\Phi$ through \eqn{g2metric}.
Hence, the condition of $G_2$ holonomy is a highly
nonlinear partial differential equation on the three-form $\Phi$.
The general strategy pursued in this paper to construct $G_2$
holonomy metrics is based on \eqn{cclosure} and \eqn{g2metric}.
This method is quite general and can naturally be
generalized to other examples than those studied in this paper.
Note that this approach was first used by \cite{bs} to construct the
first complete non-singular examples of metrics with $G_2$ holonomy
\cite{bs,gpp}.
See also \cite{hitchin} for a related approach that uses
diffeomorphism-invariant functionals of forms to study geometrical
structures in various dimensions.

 We are interested in constructing new
$G_2$ holonomy metrics on non-compact seven-manifolds.
The simplest such metrics are cones over six-manifolds
$Y$
\be\label{cone}
ds_X^2 = dr^2 + r^2 d\Omega_Y^2
\ee
where $Y$ is a compact Einstein space with weak $SU(3)$
holonomy \cite{gray}. There are three simply connected examples
known in the literature all of which are homogeneous.
The examples are
\be\label{lvlsurf}
{\bf CP}_3 = \frac{Sp(2)}{SU(2) U(1)} ~,~
F_{1,2} = \frac{SU(3)}{U(1) U(1)} ~,~
{\bf S}^3 \times {\bf S}^3 = \frac{SU(2)^3}{SU(2)} ~.
\ee
To make things more interesting we want to study
smooth deformations of these conical metrics, and
will restrict our attention to cohomogeneity one.
This means that
the level surfaces is a six dimensional space $Y$ but
the metrics on $Y$ is not Einstein anymore but depends
on the radial coordinate $r$.
Smooth deformations of the three examples which preserve
the symmetries of the conical metrics and are asymptotically
conical (AC) have already been constructed some time ago \cite{bs,gpp}.
They are the non-singular metrics on the ${\bf R^3}$ bundle over
${\bf S}^4$, the ${\bf R^3}$ bundle over ${\bf CP}^2$,
and the spin bundle
over ${\bf S}^3$ which is topologically ${\bf R}^4 \times {\bf S}^3$.

The examples that we will study in this paper are further generalizations
of the asymptotically conical metrics on ${\bf R}^4 \times {\bf S}^3$
with $SU(2)^3$ symmetry. In particular we construct metrics that
are not (AC) but only asymptotically locally conical (ALC). This
means that for large $r$ one of the directions of the manifold
does not blow up but stabilizes at a finite value. This direction is
the orbit of an $U(1)$ isometry and this requires that the isometry
group of the original (AC) metric is reduced. In compactifications of
$M$-theory on such manifolds we use this particular $U(1)$ isometry
to reduce to a vacuum solution of type IIA string theory. Because of
\eqn{m2a} the size of this $U(1)$ corresponds to the dilaton which
in contrast to the (AC) metrics is finite. Furthermore, because the
$U(1)$ has to be non-trivially fibered we will get a non-trivial RR
one-form gauge field which can be attributed to the presence of
D6-branes or RR two-form flux. In order to have D6-branes the $U(1)$
isometry has to have fix points in co-dimension four. The fix point
set has the nice interpretation as being the cycle on which the D6-brane
wraps. If there are no fix points the background is a pure flux
solution with flux over some non-trivial cycle.

In the search for new $G_2$ metrics
we take the standpoint that the central object
is the three-form $\Phi$
and the metric $g_{ij}$ is derived from it.
In the first step we make an ansatz for $\Phi$ which obeys all
the required symmetries.
Now the first equation in \eqn{cclosure} can easily be solved by
writing the three-form as the sum of a closed and an exact part
\be\label{3formans}
\Phi = \Phi_{cl} + \Phi_{ex} = \Phi_{cl} + d\Lambda
\ee
where $\Phi_{cl} \in H^3(X)$. It will turn out that choosing
a specific representative of $H^3(X)$ determines which non-trivial
cycle is blown up in $X$ to smooth out the singularity
at the tip of the corresponding conical metric.
The role of the exact piece is two-fold. It is needed to solve
for co-closure \eqn{cclosure} using \eqn{g2metric}. Furthermore,
it ensures that the three-form $\Phi$ is {\it stable} in the sense
of \cite{hitchin}. In simple terms this means that $\Phi$
has to be sufficiently generic to make the metric in
\eqn{g2metric} non-degenerate.\footnote{A small example should make
clear what we mean by this. Assume we study $X = S^3 \times R^4$, then
we could naively assume that the volume three-form on $S^3$ is a nice
harmonic three-form. But it is not generic enough and leads to
an $s_{ij}$ in \eqn{g2metric} with less than maximal rank and hence
leads to a degenerate metric.}
It is clear that the equations for co-closure impose highly
non-linear conditions and in practise they can only be worked out
using computer programs like Mathematica.
In the rest of this section we will apply our method to the case
of $SU(2)^3$ symmetric metrics on the spin bundle of ${\bf S}^3$.
This is meant as a warm-up for the new,
$SU(2)^2 U(1)$ symmetric examples of metrics that we will
study in section 3.

\subsection{Old $G_2$ Holonomy Metrics Revisited}

The metrics we want to rederive here are smooth deformations
of cones over ${\bf S}^3 \times \tilde{\bf S}^3$ and have an
$SU(2)^3$ symmetry. It is most convenient to introduce two sets
of $SU(2)$ left-invariant one-forms $\sigma_a$ and $\Sigma_a$
by\footnote{The $T^a$ are $SU(2)$ generators with $T^a T^b = \delta_{ab}+
i \epsilon_{abc} T^c$. The triplet of one-forms $\sigma_a$ can be
extracted from the $SU(2)$ valued $\sigma$ using
$\sigma^a = -(i/2) \mathrm{Tr} T^a \sigma$. This relation will be used
at various occasions throughout the paper.}
\be\label{oneforms}
U^{-1} dU = T^a \sigma_a \equiv \sigma ~,~ V^{-1} dV = T^a \Sigma_a
\equiv \Sigma
\ee
which makes an $SU(2)^2 = SU(2)_L \times \widetilde{SU(2)}_L$
symmetry manifest.
The $SU(2)$ valued matrices $U$ and $V$ parametrize the two three-spheres.
The third $SU(2)$ symmetry acts by the diagonal subgroup of the right action
\be\label{su2diag}
SU(2)_R^\mathrm{diag} =
\left( SU(2)_R \times \widetilde{SU(2)}_R \right)_\mathrm{diag} ~.
\ee
The two three-spheres can be parametrized by two independent sets of
Euler angles in terms of which the two sets of one-forms associated
with the symmetries mentioned above become
\ba
\begin{array}{lcl}
\sigma_1= \cos \psi\, d\theta + \sin \psi\, \sin \theta\, d\phi &
\quad , \quad &
\Sigma_1= \cos {\tilde\psi}\, d{\tilde\theta} + \sin {\tilde\psi}\, \sin
{\tilde\theta}\, d{\tilde\phi} \\
\sigma_2= -\sin \psi\, d\theta + \cos \psi\, \sin \theta\, d\phi &
\quad , \quad &
\Sigma_2= -\sin {\tilde\psi}\, d{\tilde\theta} + \cos {\tilde\psi}\, \sin
{\tilde\theta}\, d{\tilde\phi} \\
\sigma_3= d\psi + \cos \theta\, d \phi &
\quad , \quad &
\Sigma_3= d{\tilde\psi} + \cos {\tilde\theta}\, d{\tilde\phi}
\label{sigmas}
\end{array}
\ea
which, furthermore, satisfy the $SU(2)$ algebras
\be\label{sigmaalg}
d\sigma_1 = - \sigma_2 \wedge \sigma_3~+ \mathrm{cyclic~ perms.},  \quad
d\Sigma_1 = - \Sigma_2 \wedge \Sigma_3~+ \mathrm{cyclic~ perms.}
\ee

The next step is to construct an ansatz for the 3-form $\Phi$ which
is invariant under the symmetries
\be\label{genansatz}
\Phi = \Phi_{cl} + d \Lambda  ~.
\ee
It contains a closed piece $\Phi_{cl} \in H^3(X)$ and
an exact piece, which guarantees closure of the 3-form and the exact
piece will be fixed by imposing co-closure.
In the case at hand $X = {\bf R} \times {\bf S}^3 \times \tilde{\bf S}^3$ and
$H^3(X) = {\bf Z} \oplus {\bf Z}$. The closed piece of the 3-form
is given by a linear combination of the two volume forms of the 3-spheres
\be\label{closed}
\Phi_{cl} = r_0^3 \left( p \, \sigma_1 \wedge \sigma_2 \wedge \sigma_3 +
q \, \Sigma_1 \wedge \Sigma_2 \wedge \Sigma_3 \right) ~,
\ee
where $(p,q)$ are integers and label different elements of $H^3(X)$.
In general the metric on $X$
is a smooth deformation of the cone metric over
${\bf S}^3 \times \tilde{\bf S}^3$.
To be more specific it is the spin bundle over ${\bf S}^3$
which has topology ${\bf S}^3 \times {\bf R}^4$.
The zero section of the total bundle
space is a three-sphere with radius proportional to $r_0$.
Hence, in the limit $r_0 \to 0$ we obtain the singular cone metric
and the three-form $\Phi$ becomes exact.

The exact piece of the three-form consistent with the symmetries is
\be\label{exact}
d \Lambda = d \left( a(r) \, \sigma_a \wedge \Sigma_a \right) ~.
\ee
For this highly symmetric case imposing co-closure of the three-form
\be\label{coclosure}
d \ast_\Phi \Phi = 0 ~
\ee
does not impose any additional constraints
on $a(r)$ which is consistent with the fact that we
have not fixed reparametrization invariance yet.
Using \eqn{g2metric} we can present the most general
metric which depends on the choice of a representative
of the third cohomology class labelled by $(p,q)$,
the size of the blown up cycle $r_0$ and an
arbitrary function $a(r)$
\ba\label{phimetric}
ds^2 & = &
\Big[ a (a - p r_0^3) \, (\sigma_a)^2 +
a (a + q r_0^3) \, (\Sigma_a)^2 \nonumber  \\
 & & ~~ - (p q r_0^6 + a^2) \, \sigma_a \Sigma_a +
(a^\prime)^2 dr^2 \Big]/\Omega \\
\Omega & = & 2^{-\frac{2}{3}} \left( 3 a^4 - 4 (p-q) r_0^3 a^3 -
6 p q r_0^6 a^2 - p^2 q^2 r_0^{12} \right)^\frac{1}{3} ~.
\ea

We can choose the arbitrary function to be linear
$a(r) \equiv r$ using the reparametrization invariance of $r$
\ba\label{genmetric}
ds^2 & = &
\Big[ r (r - p r_0^3) \, (\sigma_a)^2 +
r (r + q r_0^3) \, (\Sigma_a)^2 \nonumber \\
& & ~~ - (r^2 + p q r_0^6) \, \sigma_a \Sigma_a + dr^2 \Big]/\Omega \\
\Omega & = & 2^{-\frac{2}{3}} \left( 3 r^4 - 4 (p-q) r_0^3 r^3 -
6 p q r_0^6 r^2 - p^2 q^2 r_0^{12} \right)^\frac{1}{3} ~.
\ea
It is easy to see that the metric \eqn{genmetric} is
in general singular in the interior and we want to get constraints on the
parameters $(p,q)$ by requiring the metric to be regular.
A singularity appears when $\Omega$ vanishes
because of the fractional power of the warp factor $\Omega^{-1}$.
The only situation when we can get
smooth solutions occur when $\Omega$ vanishes linearly.
For this to happen the forth order polynomial
\be\label{poly}
P = 3 r^4 - 4 (p-q) r_0^3 r^3 - 6 p q r_0^6 r^2 - p^2 q^2 r_0^{12}
\ee
must have a third order zero. Inspection of the discriminant
\be\label{disc}
\Delta \sim p^4 q^4 (p+q)^4
\ee
reveals that there are only three possible solutions
namely $(p,q) = (-1,0)$, $(1,-1)$ or $(0,1)$ up to overall
signs. Indeed only metrics corresponding to these three combinations
lead to smooth solutions and
all other combinations have singularities. (See \cite{cglp} for a
different derivation of this result.)

For large $r$ the metrics are conical which can be made
manifest if we choose a different coordinate in which
(for large $r$) $a(r) \propto r^3$.
Adding a suitable constant we can bring the $(-1,0)$ solution
to the form as it is presented in \cite{gpp}.
We find
\be\label{oldsol}
a(r) = \frac{4}{3}(r^3 - r_0^3)  ~.
\ee
The $(-1,0)$ solution
corresponds to the well known metric of $G_2$ holonomy
on the spin bundle of ${\bf S}^3$ \cite{bs,gpp}
\be\label{symmetric1}
ds^2 = 12 dr^2/\left( 1 - \frac{r_0^3}{r^3}\right) +
       r^2 \left( \sigma_a \right)^2 +
       \frac{r^2}{3} \left( 1 - \frac{r_0^3}{r^3}\right) \left(2 \Sigma_a -
                  \sigma_a \right)^2 ~.
\ee
The $(0,1)$ solution is obtained
by a simple exchange of the left invariant 1-forms
$\sigma_a \leftrightarrow \Sigma_a$ in the metric
\eqn{symmetric1}.
Finally we come to the third metric,
which was also found in \cite{bggg,cglp,atiwit},
and which corresponds to the $(1,-1)$ solution.
In this case the ${\bf Z}_2$ flip
$\sigma_a \leftrightarrow \Sigma_a$,
which exchanges the $(-1,0)$ with the $(0,1)$
solution, becomes a symmetry. It leaves the metric
invariant but transforms the three-form $\Phi \to -\Phi$.
The corresponding metric is given by
\be\label{symmetric2}
ds^2 = 12 dr^2/\left( 1 - \frac{r_0^3}{r^3}\right) +
       r^2 \left( \sigma_a - \Sigma_a \right)^2 +
       \frac{r^2}{3} \left( 1 - \frac{r_0^3}{r^3}\right) \left(\sigma_a +
                  \Sigma_a \right)^2 ~.
\ee
and $a(r) = (4 r^3 - r_0^3)/3$.

As shown in \cite{atiwit} the existence of three solutions
can be nicely explained by the triality symmetry of the conical
metric $r_0=0$. This symmetry is broken down to a residual
${\bf Z}_2$
symmetry by blowing up one of three possible three-spheres.
The broken part of the triality group permutes the three solutions, and
the action of the unbroken ${\bf Z}_2$ symmetry on the one-forms is
$\sigma \to -\sigma~,~\Sigma \to \Sigma - \sigma$ for the $(-1,0)$ solution,
$\Sigma \to -\Sigma~,~\sigma \to \sigma - \Sigma$ for the $(0,1)$ solution,
and $\sigma \leftrightarrow \Sigma$ for the $(1,-1)$ solution.

\section{New $G_2$ Holonomy Metrics}

The goal of this section is to construct generalizations of the
(AC) metrics \eqn{symmetric1} and \eqn{symmetric2}
using the the method outlined
in the previous section. In particular we are interested in metrics which
are less symmetric and are not asymptotically conical. We search
for metrics that can be reduced to type IIA solutions with finite
string coupling. This requires the existence of a $U(1)$ isometry
whose orbits have finite radius for large radius $r$.
As explained in the previous section such backgrounds correspond
to the $M$-theory lift of type IIA compactifications on Calabi-Yau
manifolds involving wrapped
branes or RR two-form flux. Note that after taking into account
the backreaction of the branes or flux the six-manifold is not
Calabi-Yau anymore.

In particular, we will study in turn three cases that correspond to the
$M$-theory lift of
\begin{itemize}
\item D6 branes wrapped on the deformed conifold
\item The resolved conifold with RR two-form flux on the blown up ${\bf S}^2$
\item The line bundle over ${\bf S}^2 \times {\bf S}^2$ with RR two-form
      flux on the  blown up ${\bf S}^2 \times {\bf S}^2$
\end{itemize}

Topologically the first two examples are ${\bf S}^3 \times {\bf R^4}$,
whereas the third example \cite{cvnew} corresponds to the
${\bf R}^2$ bundle over $T^{11}$.

\subsection{Wrapped D6 branes}

A first example of that type was found recently \cite{bggg}
and we will review
it here although we will derive it in a different fashion.
It corresponds to the uplift of a background of D6 branes
wrapped on the three-sphere inside the deformed conifold
${\bf T^* S^3}$ at finite string coupling.
As before we choose to work with the basis of one-forms
$\sigma_a ~,~ \Sigma_a$ defined
in \eqn{oneforms}.
Then the continuous symmetry of the deformed conifold
which is not altered by the presence of the D6 branes is
\be\label{symmconi}
SU(2)_L \times \widetilde{SU(2)}_L ~.
\ee
In addition there is an extra $U(1)$ symmetry corresponding to the
M-theory circle which in our basis is implemented as
\be\label{u1diag}
U(1)_R^{\mathrm{diag}} = \left(U(1)_R \times \widetilde{U(1)}_R
\right)_\mathrm{diag} ~,
\ee
which corresponds to the Killing vector
$\partial_\psi+\partial_{\tilde{\psi}}$.
The $U(1)_R^{\mathrm{diag}}$ symmetry is a left-over of the
diagonal right $SU(2)$ symmetry \eqn{su2diag}.
It acts as an $SO(2)$ rotation
simultaneously on $\sigma_a ~,~ \Sigma_a~,~a=1,2$
but leaves $\sigma_3 ~,~ \Sigma_3$ unchanged.

At this stage we can present the most general ansatz for the 3-form $\Phi$
invariant under the $SU(2)_L \times \widetilde{SU(2)}_L \times
U(1)_R^{\mathrm{diag}}$ symmetry. It has the form
\ba\label{3formansatz}
\Phi & = & r_0^3 \left( p \, \sigma_1 \wedge \sigma_2 \wedge \sigma_3 +
q \, \Sigma_1 \wedge \Sigma_2 \wedge \Sigma_3 \right) \nonumber \\
 & & \quad + \, d \left( a(r) (\sigma_1 \wedge \Sigma_1 +
\sigma_2 \wedge \Sigma_2) + b(r) \sigma_3 \wedge \Sigma_3 \right) ~.
\ea
A few comments are in order.
To write down the three-form we have made a particular
choice for $\sigma_i~,~\Sigma_i~,~i=1,2$ since we could replace them
by $\sigma^\prime_i = M_i^j \sigma_j$ and $\Sigma^\prime_i = N_i^j \Sigma_j$
without changing the equation for $a(r)$ and $b(r)$,
where $M_i^j$ and $N_i^j$ are independent, constant $SO(2)$ matrices.
This allows us to get rid of terms of the form
$d \left( f(r) (\sigma_1 \wedge \Sigma_2 - \sigma_2 \wedge \Sigma_1) \right)$.
Furthermore, we have excluded terms of the form
$d \left( g(r) \sigma_1 \wedge \sigma_2 + h(r) \Sigma_1 \wedge \Sigma_2
\right)$
although they are allowed by the symmetries. The reason is that
they lead to terms in the metric that mix radial and angular directions.
With this in mind \eqn{3formansatz} is the most general
three-form ansatz that obeys the symmetries \eqn{symmconi}
and \eqn{u1diag}.

Finally, we have to impose the discrete ${\bf Z}_2$ symmetry of the
conifold which in our basis exchanges the two sets of left-invariant
one-forms $\sigma \leftrightarrow \Sigma$.
This symmetry leaves the
metric invariant but flips the sign of the three-form
$\Phi \to - \Phi$, hence we have to take
\be\label{pq}
(p,q) = (1,-1)~.
\ee

The general expression for the metric following from
\eqn{phimetric}, \eqn{3formansatz} and \eqn{pq} is
\ba\label{wD6metric}
ds^2 & = & \Big[ \frac{1}{4} (b - r_0^3)(2 a + b + r_0^3) a^\prime
\left[
(\sigma_1 - \Sigma_1)^2 + (\sigma_2 - \Sigma_2)^2 \right]  \nonumber \\
 & & + \frac{1}{4} (b - r_0^3)(2 a - b - r_0^3) a^\prime \left[
(\sigma_1 + \Sigma_1)^2 + (\sigma_2 + \Sigma_2)^2 \right]  \nonumber \\
 & & + \frac{1}{4}(4 a^2 - (b + r_0^3)^2) b^\prime (\sigma_3 - \Sigma_3)^2 \\
 & & + \frac{1}{4} (b - r_0^3)^2 b^\prime (\sigma_3 + \Sigma_3)^2 +
(a^\prime)^2 b^\prime dr^2 \Big]/\Omega \nonumber \\
\Omega & = & \frac{1}{2^{2/3}} (b - r_0^3)^{2/3}
(4 a^2 - (b + r_0^3)^2)^{1/3} (a^\prime)^{2/3}
(b^\prime)^{1/3}  ~. \nonumber
\ea
Imposing co-closure \eqn{cclosure} leads to a non-linear second order
differential equation for $a(r)$ and $b(r)$
\ba\label{diffeqn}
& & 4 a' b' ( a (r_0^3 - b) a' + (2 a^2 - b (r_0^3 + b)) b') \nonumber \\
& & + (r_0^3 - b)(r_0^3 - 2 a + b)(r_0^3 + 2 a + b) (a' b'' - a'' b') = 0~,
\ea
where we used $^\prime \equiv \frac{d}{dr}$. The differential equation
\eqn{diffeqn} is second order which seems to contradict the fact that
closure and co-closure impose first order equations on the three-form.
However, in writing the ansatz \eqn{3formansatz} we already imposed
closure thus replacing certain functions by first derivatives of others.
We also obtained only one equation in two functions, since reparametrization
invariance allows to choose one of them freely. However, this choice is not
completely arbitrary and the function has to be chosen such that
the metric \eqn{wD6metric} does not become degenerate.

Despite the formidable form of the equation,
it is possible to find solutions in special cases.
In general we expect a two parameter family of non-singular
solutions once one of the functions is fixed by reparametrizations.
In type IIA these two parameters correspond to the asymptotic value
of the dilaton and the size of the blown up three-sphere in the
deformed conifold.

The simplest solution was already discussed in the previous section which
corresponds to the $SU(2)^3$ symmetric $(1,-1)$ solution
\eqn{symmetric2} with
\be\label{soln01}
a = b = (4 r^3 - r_0^3)/3 ~.
\ee

Things become more interesting when the two functions are not equal.
A particular one parameter solution can be found if we assume
$a$  and $b$ to be  polynomials of finite degree
in the radial coordinate $r$. Inserting this ansatz into \eqn{diffeqn}
we find
\be\label{sol02}
a = \frac{1}{18}(r^3 - 3 r_0^2 r) ~,~
b = \frac{1}{9} (2 r_0 r^2 - 9 r_0^3) ~.
\ee
This gives the solution found in \cite{bggg}\footnote{The mismatch
of certain numerical coefficients is due to different conventions
used in this paper.}
\ba\label{bgggmetric}
ds^2 & = & A^2 \left[(\sigma_1 - \Sigma_1)^2 + (\sigma_2 - \Sigma_2)^2 \right]
+ B^2 \left[(\sigma_1 + \Sigma_1)^2 + (\sigma_2 + \Sigma_2)^2 \right]
\nonumber \\
 & & + D^2 (\sigma_3 - \Sigma_3)^2 + 4 r_0^2 C^2 (\sigma_3 + \Sigma_3)^2 +
dr^2/C^2
\ea
with
\ba\label{bgggsoln}
A(r)^2 & = & \frac{1}{12}(r - r_0) (r + 3 r_0) ~, \nonumber \\
B(r)^2 & = & \frac{1}{12}(r + r_0) (r - 3 r_0) ~,\nonumber \\
C(r)^2 & = & \frac{(r - 3 r_0)(r + 3 r_0)}{(r - r_0)(r + r_0)} ~, \nonumber \\
D(r)^2 & = & \frac{r^2}{9} ~.
\ea
The metric is complete and well-defined in the range
$r \in [ 3 r_0,\infty )$.
This metric can be thought of as a Taub-NUT space fibered over $S^3$
and the base of this fibration can be found at $r = 3 r_0$.
It is parametrized by $(U,V) = (g,g^{-1})$ with $g \in SU(2)$
and the solution is a deformation
of the $(1,-1)$ solution \eqn{symmetric2} of the previous section.
The peculiar feature of this
metric is that it does not behave conically in all directions but, as
we can see from the behaviour of $C(r)$ for large $r$, there is one
direction that stabilizes at large radii. The corresponding $U(1)$ isometry
is generated by the Killing vector $\partial_{\psi} +
\partial_{\tilde{\psi}}$.

The triality symmetry \cite{atiwit} of the conical metric predicts
that there exist similar metrics which are deformations of the $(-1,0)$
and $(0,1)$ solutions.
The metrics we will present in the following are new,
however, the differential equations and the coefficient functions
turn out to be the same, so we do not have to repeat them.

It is interesting to note that we cannot obtain these
metrics by using the ansatz \eqn{3formansatz} with
the same one-forms \eqn{sigmas} and replacing \eqn{pq} by
$(-1,0)$ or $(0,1)$. However, such ans\"atze will play a
vital role in the next subsection when we
construct M-theory uplifts of the type IIA background of the
resolved conifold with RR two-form flux.

The new solutions can be expressed in terms of the known one
\eqn{bgggmetric} by replacing the two sets of
left-invariant one-forms \eqn{sigmas} in an
appropriate fashion. Before presenting the result let us
explain the reason behind these replacements.
The level surfaces of our metrics are
homogeneous spaces of the form
\be\label{levelsurface}
\frac{SU(2) \times SU(2) \times SU(2)}{SU(2)_D} \sim
{\bf S}^3 \times {\tilde{\bf S}}^3 ~.
\ee
By virtue of the coset structure there is an $SU(2)^3$ symmetry
which acts by left multiplication on each of the three $SU(2)$
factors. Furthermore, there is an triality symmetry which acts
by permuting the three $SU(2)$ factors \cite{atiwit}.
If we want to deform the conical metric we can blow up one of
three three-spheres corresponding to one of three $SU(2)$ factors.
In homology these three three-spheres are not independent and
they obey a linear relation $D_1+D_2+D_3=0$,
where the $D_i~,~i=1,2,3$ denote the homology classes
of the three-spheres. The coset \eqn{levelsurface}
can be represented in homogeneous coordinates by three
$SU(2)$ elements $a,b,c$ modulo the identification
$(a,b,c) \sim (a \lambda, b \lambda, c \lambda)$ where
$\lambda \in SU(2)$.

Now by choosing the particular patch $c = \lambda^{-1}$
we can match this to the parametrization we chose in the
definition of our one-forms
\eqn{oneforms} in terms of $U, V$
\be\label{match}
(a c^{-1},b c^{-1}, 1) = (U, V, 1) ~.
\ee
In this parametrization the $(-1,0)$ sphere is parametrized by
$(U,V) = (g^{-1},1)$,
the $(0,1)$ sphere by $(U,V) = (1,g)$, and the $(1,-1)$ sphere by
$(U,V) = (g,g^{-1})$ with $g~\in~SU(2)$.
Our solution \eqn{bgggmetric} corresponds to $(1,-1)$,
how do we get the other ones?

The answer is
\ba
\sigma \equiv T^a \sigma_a = U^{-1} dU & \to &
V (\sigma - \Sigma) V^{-1} = \left( U V^{-1} \right)^{-1}
d \left( U V^{-1} \right) \nonumber \\
\Sigma \equiv T^a \Sigma_a = V^{-1} dV & \to & - V \Sigma V^{-1} =
V d \left( V^{-1} \right)
\ea
which can be understand by going to a different patch
$(U, V, 1) \to (U V^{-1}, 1, V^{-1})$.
For the sums and differences that appear in the metric
this means
\ba\label{repl1}
\sigma - \Sigma & \to & V \sigma V^{-1} \nonumber \\
\sigma + \Sigma & \to & V (\sigma - 2 \Sigma) V^{-1}
\ea
and, therefore, in this case we find at $r = 3 r_0$ a finite
size three-sphere parametrized by $U$ i.e. the $(-1,0)$ three-sphere.
Compared to the $SU(2)^3$ symmetric case \eqn{symmetric1}
the one-forms are
rotated non-trivially. But
in the $SU(2)^3$ symmetric case these rotation would never show up
in the expressions for the metric and $\Phi$.

Finally, the $(0,1)$ case can be obtained in the same fashion by going to
the patch $(U, V, 1) \to (1, V U^{-1}, U^{-1})$. This translates
into replacing
\ba\label{repl2}
\sigma - \Sigma & \to & U \Sigma U^{-1} \nonumber \\
\sigma + \Sigma & \to & U (\Sigma - 2 \sigma) U^{-1}
\ea
which is just a $U \leftrightarrow V$ flip of the solution
obtained from \eqn{bgggmetric} using \eqn{repl1}.

Therefore, we are able to identify three solutions
that are fibrations of a Kaluza-Klein monopole over
a three-sphere which agrees with expectations from
triality. The metrics are smooth and complete.
They can be used to describe four-dimensional vacua
with $\mathcal{N}=1$ supersymmetry of the type
$R^4 \times X$ where $X$ is any of the three metrics.
If we reduce this solution along the particular $U(1)$
isometry to type IIA we get a background that describes a D6 brane
wrapped on the three-sphere inside the deformed
conifold. In this case the dilaton interpolates between zero
and a finite value set by $r_0$ as $r$ varies from $r = 3 r_0$ to
infinity. The string frame metric has small curvature over most
of the manifold but blows up over the three-sphere at $r = 3 r_0$
because the dilaton
$e^\phi = r_0^{3/4} C^{3/2}$ vanishes there \cite{bggg}.

It is straightforward to generalize these metrics so that they
describe a stack of $N$ wrapped D6 branes.
For this we have to mod out the the $G_2$ holonomy metric by
a ${\bf Z}_N$ action as in \cite{bggg}.
In this case also the M-theory solution is singular, there is a
${\bf R}^4/{\bf Z}_N$ orbifold singularity at $r = 3 r_0$, which
gives rise to $SU(N)$ enhanced gauge symmetry. At low energies
we find ${\cal N} = 1$ SYM with $SU(N)$ gauge group in four
dimensions coupled to eleven-dimensional supergravity.
For more details on the  metric \eqn{bgggmetric}
and its reduction to type IIA we refer
the reader to \cite{bggg}. (The $(-1,0)$ and $(0,1)$ metrics have
the same properties.)


\subsection{Resolved conifold with RR two-form flux}

We continue our search for new $G_2$ metrics
and study the three-form ansatz \eqn{3formansatz} with
\be\label{pqflux}
(p,q) = (-1,0)~.
\ee
This will lead us to the construction of a
novel class of metrics that correspond to the M-theory
uplift of type IIA backgrounds of the resolved conifold
metric with RR two-form flux turned on over the blown up ${\bf S}^2$.
From \eqn{3formansatz} and \eqn{pqflux} we obtain the metric
\ba\label{fluxmetric}
ds^2 & = & A^2 \left[ (\sigma_1)^2 + (\sigma_2)^2 \right]
         + B^2 (\sigma_3)^2  \nonumber \\
 & & + C^2 \left[ \left(\Sigma_1 - f \sigma_1
\right)^2 +
\left( \Sigma_2 - f \sigma_2 \right)^2 \right] + D^2
\left( \Sigma_3 - g \sigma_3
\right)^2 + E^2 dr^2
\ea
with
\ba
A^2 & = & \frac{a^\prime}{4 a} (4 a^2 (b + r_0^3) - b^3)/\Omega ~,~
B^2 ~ = ~ \frac{b b^\prime}{4 a^2}(4 a^2 (b + r_0^3) - b^3)/\Omega
~, \nonumber \\
C^2 & = & a b a^\prime/\Omega ~,~ D^2  ~ = ~ a^2 b^\prime/\Omega ~,~
E^2 ~ = ~ (a^\prime)^2 b^\prime/\Omega ~, \nonumber \\
f & = & \frac{b}{2a} ~,~ g ~ = ~ 1-2 f^2~, \nonumber \\
\Omega & = & \frac{1}{2^{2/3}} b^{1/3}
(4 a^2 (b + r_0^3) - b^3)^{1/3} (a^\prime)^{2/3} (b^\prime)^{1/3} ~,
\ea
and a second order differential equation for $a(r)$ and $b(r)$
\ba\label{diffeqn1}
& & 4 a' b' \left[ a b (b + r_0^3) a' + (b^3-a^2 (r_0^3 + 2 b)) b'
\right] \nonumber \\
& & + b (b^3-4 a^2 (r_0^3 + b)) (a' b'' - a'' b') = 0 ~.
\ea

A particular solution of this rather complicated equation \eqn{diffeqn1}
was already presented in section 2. It corresponds to the $(1,0)$
solution \eqn{symmetric1} with
\be\label{sol03}
a = b = \frac{4}{3} (r^3 - r_0^3) ~.
\ee
A second analytic solution can be obtained by assuming that $a$ and $b$
are polynomials of finite degree in $r$. One solution that can
be found easily is
\be\label{singflux}
a = r^3 ~,~ b = -\frac{2 r_0}{7^{1/3}} r^2 ~,
\ee
however, the corresponding metric turns out to be singular at
$r=0$. Therefore,
we will not discuss it further and move on to study numerical
solutions.

In order to study \eqn{diffeqn1} numerically we solve it perturbatively
in the interior which we choose to be at $r=0$ and integrate
numerically to large radii to obtain the asymptotic behaviour.
We will compare
these numerically obtained asymptotics with a perturbative solution
at $r=\infty$.
For convenience we choose $a = r^3$ and with this input we
study \eqn{diffeqn1}
perturbatively around $r = 0$.
At this point it is important to understand
which boundary conditions we have to impose and
this requires knowledge of the geometry in the interior.
First of all we do not want the finite size circle to shrink
to zero at $r=0$ because we seek a solution that corresponds in
type IIA to a background with no D6 branes and only flux.
This means that the circle we use to reduce to type IIA must not lie
in the fiber direction but has to mix with the base direction of the
spin bundle over ${\bf S}^3$ \cite{amv}.
The circle is again generated by the Killing vector $\partial_\psi +
\partial_{\tilde{\psi}}$. Hence the size of the circle is given by
\be\label{size}
B^2 + (1-g)^2 D^2
\ee
which means that $B^2$ has to remain finite, since $D^2$ being part
of the fiber directions vanishes in the interior.
The coefficient function $B^2$ sets the size of $U(1)$ Hopf-fiber
of the base ${\bf S}^3$ of the fibration. As we want to keep
the expression in \eqn{size} finite for all radii, $B^2$ has
to stabilize at $r \to \infty$ whereas $D^2 \propto r^2$ and
$(1-g) \propto 1/r^2$ so that $D^2 (1-g)^2 \propto 1/r^2$.
The coefficients $A^2$ and $B^2$ in general do not agree in the
interior
although they do so in the $SU(2)^3$ symmetric solution.
This means that the $U(1)$ fiber of the
base three-sphere is in general squashed.

The boundary condition at $r=0$
that guarantees that $A^2$ and $B^2$ are finite and that the fiber part
of the metric approaches the flat metric on ${\bf R}^4$ is
\be\label{bcflux}
b \sim (1 - y) r^3 + {\mathcal O}(r^6) ~.
\ee
The parameter $y$ measures the deviation from the $SU(2)^3$ symmetric
solution. As numerical studies with the boundary condition \eqn{bcflux}
show $y$ is related to the asymptotic value of the dilaton.
Note that in general we have two integration constants, one of which
is fixed by the boundary condition \eqn{bcflux} and one, $y$, is
left as a free parameter. In addition we have the parameter $r_0$
from the three-form ansatz
which sets the scale for the blown up two-sphere in the
resolved conifold. So we expect a two-parameter family of smooth
solutions.

Next we wish to solve \eqn{diffeqn1} perturbatively with the
boundary condition \eqn{bcflux} and use these solutions as starting
values for a numerical integration from $r \gtrsim 0$.
In the interior it is useful to use the coordinate $\rho \equiv r^3$.
The perturbative solution around $\rho=0$ is
\be\label{numsol}
b = (1 - y) \rho - \frac{1}{3 r_0^3} (2 y - 5 y^2 + 4 y^3 - y^4) \rho^2 +
\ldots ~.
\ee
The corresponding metric to lowest order in $\rho$ is
\ba\label{ajf}
ds^2 & \sim & \frac{1}{(1-y)^{2/3}}\Bigg\{ r_0^2 \left[ (\sigma_1)^2 +
(\sigma_2)^2 \right] +
r_0^2 (1-y)^{2} (\sigma_3)^2 \Bigg\} \nonumber \\
&& ~~~~ +\frac{(1-y)^{1/3}}{r_0} \rho \Bigg\{
\bigg( \Sigma_1 - \frac{1-y}{2} \sigma_1 \bigg)^2 +
\bigg( \Sigma_2 - \frac{1-y}{2} \sigma_2 \bigg)^2
\nonumber \\
&& ~~~~ +\left( \Sigma_3 - \frac{1 + 2 y - y^2}{2}
\sigma_3 \right)^2 + \frac{d\rho^2}{\rho^2} \Bigg\} ~,
\ea
which corresponds to a squashed three-sphere with an ${\bf R}^4$
fibered over it. Numerical integration of \eqn{diffeqn1} shows
that this perturbative solution can be smoothly extended to
$r \to \infty$ where we find $b \sim c(y) r^2$. The positive
constant $c(y)$ depends on the coefficient $y$ in the boundary condition
\eqn{bcflux}. The large $r$ behaviour of $b$ is consistent with the existence
of a one parameter family of perturbative solutions at $r \to \infty$
\ba\label{numflux2}
b & = & c(y) r^2 -\frac{7 c(y)^3 + 8 r_0^3}{16} +
\frac{81 c(y)^6 + 160 c(y)^3 r_0^3 + 64 r_0^6}{256 c(y) r^2}
\nonumber \\
& & - \frac{9 c(y) (63 c(y)^6 + 128 c(y)^3 r_0^3 + 64 r_0^6)}{1024 r^4}
\log (r) + \mathcal{O}(r^4).
\ea
Note that the asymptotic
solution contains logarithmic terms which are necessary
to obtain solutions with positive $c(y)$. If we do not allow for
log terms we can only get solutions with negative $c(y)$ e.g.
the singular solution \eqn{singflux}.

The asymptotic form of the metric for $r \to \infty$  is
\ba\label{asymflux}
ds^2 & = & \frac{r^2}{6} \left[ (\sigma_1)^2 +
(\sigma_2)^2 \right] +
\frac{c^2}{9} (\sigma_3)^2  \nonumber \\
&& + \frac{r^2}{6} \left[
\left( \Sigma_1 - \frac{c}{2 r} \sigma_1 \right)^2 +
\left( \Sigma_2 - \frac{c}{2 r} \sigma_2 \right)^2  \right]
\nonumber \\
&& +\frac{r^2}{9} \left( \Sigma_3 - (1 - \frac{c^2}{2 r^2})
\sigma_3 \right)^2 + dr^2 ~.
\ea
This is the standard conifold metric with a finite size circle
fibered over it.
The allowed range of the parameter $y$ is $0 \leq y \leq 1$.
If $y=0$ we see from the first line of \eqn{ajf} that the
three-sphere is not squashed and the full solution
corresponds to the asymptotically conical solution with
$SU(2)^3$ symmetry \eqn{symmetric1}.
For increasing $y$ the three-sphere
in the interior is more and more squashed and
the value of $c(y)$, which is related to the
asymptotic value of the dilaton, decreases.
Finally at $y=1$ the circle shrinks to zero size and the remaining
six-dimensional metric becomes the standard metric on the resolved
conifold. In order to find the metric in the limit $y \to 1$
we have to perform the following rescaling
\be
ds^2 \to ds^2/(1-y)^{2/3} ~,~ r \to r (1-y) ~,
\ee
as can be seen from \eqn{ajf}.

\underline{\it Reduction to type IIA string theory}

Here we want to provide some details on the reduction of the new $G_2$
holonomy metric \eqn{fluxmetric} to type IIA string theory
using \eqn{m2a} since this is slightly more involved in this case than
in \eqn{wD6metric}.
The relevant $U(1)$ isometry is generated
by the Killing vector $\partial_\psi + \partial_{\tilde{\psi}}$.
Having this in mind we rewrite the metric \eqn{fluxmetric} which
makes this manifest
\ba\label{flux2a}
ds_{11}^2 & = &dx_4^2+ A^2 \left[ (\sigma_1)^2 + (\sigma_2)^2 \right] +
C^2 \left[ (\Sigma_1 - f \sigma_1)^2 + (\Sigma_2 - f \sigma_2)^2\right]
\nonumber \\
& & + \frac{B^2 D^2}{B^2 + (1-g)^2 D^2}
\left( \sigma_3 - \Sigma_3 \right)^2 + E^2 dr^2
\nonumber \\
 & & + \frac{1}{4} \left[ B^2 + (1-g)^2 D^2 \right]
\left[ \sigma_3+\Sigma_3 + \frac{B^2-D^2(1-g^2)}{B^2+(1-g)^2 D^2}
\left(\sigma_3-\Sigma_3 \right) \right]^2 ~,
\ea
where $dx_4^2$ denotes flat $d=4$ Minkowski space.
Note that in this metric nothing depends of $\psi + \tilde{\psi}$.
Now Kaluza-Klein reduction simply amounts to dropping the last
line in \eqn{flux2a} which has been written as a complete square
for that purpose.
In particular we can now read off the dilaton
\be\label{fluxdil}
e^\phi = 2^{-3/2} \left[ B^2 + (1-g)^2 D^2 \right]^{3/4}~,
\ee
and the RR one-form gauge field
\be\label{fluxRR}
A_\mu dx^\mu = \frac{B^2-D^2(1-g^2)}{B^2+(1-g)^2 D^2}
\left(\sigma_3-\Sigma_3 \right) + \cos \theta d\phi +
\cos \tilde{\theta} d\tilde{\phi} ~.
\ee
The ten-dimensional metric in string frame is given by
\ba\label{IIaflux}
ds^2_{IIA} & = & \frac{1}{2}
\Big\{ dx_4^2 +
A^2 \left[ (\sigma_1)^2 + (\sigma_2)^2 \right] +
C^2 \left[ (\Sigma_1 - f \sigma_1)^2 + (\Sigma_2 - f \sigma_2)^2\right]
\nonumber \\
& & + \frac{B^2 D^2}{B^2 + (1-g)^2 D^2}
\left( \sigma_3 - \Sigma_3 \right)^2 + E^2 dr^2 \Big\} \times
\left[ B^2 + (1-g)^2 D^2 \right]^{1/2}
\ea
From the asymptotic solution \eqn{asymflux} we can read off the
asymptotic value of the dilaton and measure the RR two-form flux
at infinity
\be
e^\phi_\infty = \left( \frac{c(y)}{6} \right)^{3/2} ~,~
F_\infty = d A_\infty = \sin \theta d\phi \wedge d\theta + \ldots
\ee
where the dots denote terms that do not contribute to the integral
of the two-form over the blown-up two-sphere which is parametrized
by $\phi,\theta$. Hence, we get precisely one unit of RR two-form flux
over the two-sphere. Since the warp factor in
\eqn{IIaflux} is finite everywhere the type IIA metric is
completely non-singular.
It is straightforward to generalize this metric to describe in ten
dimensions a metric with $N$ units of RR flux. We simply
have to mod the original eleven-dimensional metric by a ${\bf Z}_N$
action that acts in the direction of the $U(1)$ isometry as
\be
{\bf Z}_N: (\psi,\tilde{\psi}) \to
(\psi + \frac{4 \pi}{N},\tilde{\psi}+ \frac{4 \pi}{N}) ~.
\ee
In contrast to the metrics of section 3.1 the $U(1)$ action has
no fixed points. Therefore, the ${\bf Z}_N$ orbifold of
the eleven-dimensional metric and the corresponding type IIA solution
are perfectly smooth. They provide (finite string coupling version) of
smooth supergravity duals of four-dimensional $\mathcal{N}=1$ SYM at
strong coupling \cite{ach1,amv,ach2,atiwit}.

\underline{More $G_2$ metrics}

By similar transformations as in section 3.1 we can construct five
more $G_2$ holonomy metrics from \eqn{fluxmetric}.
First of all the ${\bf Z}_2$ flip $\sigma \leftrightarrow \Sigma$,
which is a symmetry of the $(1,-1)$ solution,
generates another solution that corresponds to $(p,q) = (0,1)$.
In type IIA string theory this new metric is related to \eqn{IIaflux}
by a flop transition \cite{atiwit}.

Furthermore, we can construct from the $(-1,0)$ and $(0,1)$ solution
four more solutions using the replacements
\eqn{repl1} and \eqn{repl2}
\ba
\sigma & \to & V (\sigma-\Sigma) V^{-1} ~,~ \nonumber \\
\Sigma & \to & -V \Sigma V^{-1}~,
\ea
and
\ba
\sigma & \to & -U \sigma U^{-1} ~,~ \nonumber \\
\Sigma & \to & U (\Sigma-\sigma) U^{-1}~.
\ea

In total we obtain three metrics (from section 3.1) that correspond
to wrapped D6 branes and six metrics that correspond to the resolved
conifold with RR two-form flux. Under triality they naturally can
be arranged into three groups with three metrics each.
Each group is
distinguished by a unique $U(1)$ isometry with finite orbit and
the three different metrics in each group correspond to blow-ups of
one of the three possible spheres. In each group there is one
deformed conifold and two resolved conifolds which are related by
the familiar flop and conifold transitions in string theory.


\subsection{Flux on the line bundle over ${\bf S}^2 \times {\bf S}^2$}

In \cite{cvnew} evidence was found that for metrics of the type
\eqn{wD6metric} there exists a new branch of solutions for
the $G_2$ holonomy conditions that has a quite different
geometry than the metrics we discussed in section 3.1.
On that branch the behaviour in the interior of the space is such that
only one circle direction shrinks to zero size.
The geometry of the space is that of the $R^2$ bundle over $T^{11}$.
In order to make the metric
nonsingular in the interior the periodicity of $\psi + \tilde{\psi}$
has to be changed from its original value of $4 \pi$ to $2 \pi$,
so that asymptotically
the metric becomes a $U(1)$ fibration over the line bundle
over ${\bf S}^2 \times {\bf S}^2$ \cite{pagepope}
which is asymptotic
to a cone over $T^{11}/{\bf Z}_2$.
The particular feature of this solution is that the size of the
$U(1)$ fiber is always finite similar to the $U(1)$ isometry
of the metrics we found in section 3.2.
Note that the fibration is non-trivial and
this metric is not a simple product manifold
of a circle times a six-dimensional manifold.
This would contradict the fact that this space carries
a metric with $G_2$ holonomy.
Since the $U(1)$ action
has no fixed points the corresponding type IIA background does
not involve D6-branes. It corresponds to a compactification
on the line bundle over ${\bf S}^2 \times {\bf S}^2$ \cite{pagepope}
with
RR two-form flux over the blown-up ${\bf S}^2 \times {\bf S}^2$
cycle.

Here we would like to present some more numerical evidence that these
solutions exist by analyzing eqn. \eqn{diffeqn} perturbatively
in the interior.
Once, we have identified the local solution
we integrate \eqn{diffeqn} numerically
using the perturbative solution to
obtain initial values. This provides us with
an asymptotic solution that we can compare with
perturbative solutions around $r = \infty$.
In the following we will parametrize the metric as
\ba\label{t11metric}
ds^2 & = & A^2 \left[(\sigma_1 - \Sigma_1)^2 + (\sigma_2 - \Sigma_2)^2 \right]
+ B^2 \left[(\sigma_1 + \Sigma_1)^2 + (\sigma_2 + \Sigma_2)^2 \right]
\nonumber \\
 & & + D^2 (\sigma_3 - \Sigma_3)^2 + C^2 (\sigma_3 + \Sigma_3)^2 +
E^2 dr^2 ~,
\ea
where the relation of the coefficient functions $A,B,C,D,E$
to $a$ and $b$ can be read off from \eqn{wD6metric}.
At this point it is convenient to perform a coordinate transformation
on the angular coordinates as in \cite{bggg}
which allows us to rewrite the metric as
\ba\label{t11coord}
ds^2 & = & A^2 \left[(g_1)^2 + (g_2)^2 \right]
+ B^2 \left[(g_3)^2 + (g_4)^2 \right] + C^2 (g_6)^2 \nonumber \\
 & & + D^2 (g_5)^2 + E^2 dr^2
\ea
with
\ba\label{t11transf}
g_1 & = & -\sin \theta_1 d\phi_1 - \cos \psi_1 \sin \theta_2 d\phi_2 +
          \sin \psi_1 d\theta_2           \nonumber \\
g_2 & = & d\theta_1 - \sin \psi_1 \sin \theta_2 d\phi_2 -
          \cos \psi_1 d\theta_2           \nonumber \\
g_3 & = & -\sin \theta_1 d\phi_1 + \cos \psi_1 \sin \theta_2 d\phi_2 -
          \sin \psi_1 d\theta_2           \nonumber \\
g_4 & = & d\theta_1 + \sin \psi_1 \sin \theta_2 d\phi_2 +
          \cos \psi_1 d\theta_2           \nonumber \\
g_5 & = & d\psi_1 + \cos \theta_1 d\phi_1 +
                    \cos \theta_2 d\phi_2 \nonumber \\
g_6 & = & d\psi_2 + \cos \theta_1 d\phi_1 -
                    \cos \theta_2 d\phi_2
\ea
where the $\psi_1$ and $\psi_2$ coordinates are $4 \pi$ periodic for
the metrics studied in section 3.1 and \cite{bggg}.
We will see in a moment that this has to be modified.

As a first step we set $a = r^3$ and solve \eqn{diffeqn}
perturbatively for $b$ around $r=0$.
For $r \to 0$ we impose the boundary conditions
\be\label{bc}
b = -r_0^3 + \mathrm{const.} \times r^6
\ee
which ensures that only $D^2$ and $E^2$ vanish.
Since $A^2$, $B^2$ and $C^2$ remain finite we can see that
the first line of \eqn{t11metric} corresponds to a
(in general non-Einstein) metric on $T^{11}$.
The power series solution we find around $r=0$ is
\be\label{bpertt11}
b \sim -r_0^3 - y r^6 + \frac{y^2 (8 - 5 r_0^3 y)}{16 r_0^3} r^{12} -
\frac{y^3(80 - 88 r_0^3 y + 35 r_0^6 y^2)}{192 r_0^6} r^{18} + \ldots ~.
\ee
It is convenient to write the metric near $r=0$ in a new
coordinate $\rho \equiv r^3$. To lowest order in $\rho$ we find
\ba\label{intt11metric}
ds^2 & \sim & \frac{1}{y^{1/3}} \Big\{ \frac{r_0}{2} \left[
(g_1)^2 + (g_2)^2 + (g_3)^2 + (g_4)^2 \right] + r_0^4 y (g_6)^2 \nonumber \\
 & & ~~~ + \frac{y}{r_0^2} \left[ d\rho^2 + \rho^2 (g_5)^2 \right] \Big\}
\ea
The first line of this metric corresponds to
five dimensional manifold which is a $S^1$ bundle over $S^2 \times S^2$.
This space is also called $T^{11}$ and carries a non-Einstein metric.
The second line of the metric becomes the flat metric on ${\bf R}^2$ if
we change the periodicity of $\psi_1$ from $4 \pi$ to $2 \pi$.
This is the reason behind the change of periodicity and
gives rise to the change in the asymptotic behaviour \cite{cvnew}.

Since we were not able to find analytic solutions for the boundary
conditions \eqn{bc} we have to work numerically.
For this we use the perturbative solution \eqn{bpertt11} as
initial conditions for a numerical integration starting from
$r = \delta \gtrsim 0$.

Numerical investigations reveal that only solutions in a particular
parameter regime $0 \leq y \leq y_\mathrm{max}$ are non-singular and
extend to $r = \infty$. For large $r$ we find
numerically $b \sim - c(y) r^2$ where $c(y)$
is a positive constant that increases with increasing $y$.
This matches nicely onto perturbative solutions for large $r$
\ba
b & = & -c(y) r^2 + \frac{7 c(y)^3 + 16 r_0^3}{16} -
\frac{c(y)^2 (91 c(y)^3 + 320 r_0^3)}{256 r^2} \nonumber \\
& & + \frac{c(y) (567 c(y)^6 + 2304 c(y)^3 r_0^3 + 1024 r_0^6)}{1024 r^4}
\log(r) + \mathcal{O}(r^4) ~,
\ea
where $c(y)$ is a $y$ dependent positive constant. Note also the
presence of logarithmic terms. We denoted only the first term with
a log but at higher order in the expansion terms with higher powers
of logs appear.
Asymptotically, we get the following metric
\be\label{asymsol}
ds^2 \sim 3 \, 6^{2/3} \bigg[ \frac{r^2}{12}
\big[ (g_1)^2 + (g_2)^2 +(g_3)^2 +(g_4)^2) \big] + \frac{r^2}{9}
(g_5)^2 + \frac{c(y)^2}{36} (g_6)^2 + dr^2 \bigg]
\ee
which corresponds to the line bundle over ${\bf S}^2 \times {\bf S}^2$
\cite{pagepope} with a finite size
circle parametrized by $g_6$ fibered over it non-trivially.
The metric on the line bundle over ${\bf S}^2 \times {\bf S}^2$
is asymptotic to a ${\bf Z}_2$ orbifold of the conifold.
The interesting feature of this metric is the finite size circle
that does not shrink to zero anywhere like the metrics found in
section 3.2.
The corresponding $U(1)$ action that acts by shifts of $\psi_2$
has no fixed points and therefore the type IIA background
we obtain by Kaluza-Klein reduction on the circle has no D6-branes
that source the flux. It is a solution with pure RR-twoform flux over
the base of the line bundle over ${\bf S}^2 \times {\bf S}^2$
which is given by two two-spheres parametrized
by $\theta_i ,~ \phi_i ;~ i=1,2$. From the explicit form of
$g_6$ \eqn{t11transf} we see that the charges over the two-spheres
are $+1$ and $-1$ respectively. Note that the type IIA background
is completely non-singular, even if we increase the number
of units of flux
$(1,-1) \to (N,-N)$ by changing the periodicity of $\psi_2$ to
\be
\psi_2 \sim \psi_2 + 4 \pi/N.
\ee

The function $c(y)$ is related to the asymptotic value of the
dilaton in the type IIA solution and grows monotonically with $y$.
So in the range $0 \leq y \le y_{max}$ the metric is asymptotically
locally conical (ALC), however for the borderline case $y = y_{max}$
the metric becomes asymptotically conical. The numerical value for
the limiting value is $y_{max} \sim .42$ and agrees with the
result found in \cite{cvnew} for which we have to identify $y_{max}$
with $q_0$ of \cite{cvnew} as $y_{max} = (q_0)^2/2$.

\section{Discussion}

In this paper we made progress in the construction of new
$G_2$ holonomy manifolds using directly the definition
of torsion-free $G_2$ structures in terms of closed and
co-closed three-forms. This approach avoids shortcomings
of other approaches that start from an ansatz for the metric
since it reduces the symmetry group of the tangent space
directly to $G_2$.
The examples we study in detail are $G_2$ structures on
the space $X = {\bf S}^3 \times {\bf R}^4$, but it should be
stressed that this method can be generalized to other cases.
We generalize the
previously known metrics that are conical or asymptotically
conical (AC) \cite{bs, gpp}. By constructing a general ansatz
for the three-form we derive the general conditions for
$G_2$ holonomy with $SU(2) \times SU(2) \times U(1)$ symmetry.
The general form of the three-form, metric and the condition
for $G_2$ holonomy in form of a second order differential
equation are summarized in appendix A.
We study three classes of solutions in some detail.
The one example of the first class was found recently \cite{bggg}
using a less general method and can be thought of as
a fibration of the KK monopole over ${\bf S}^3$. In fact there
exist two more examples which correspond to the blow up of one
of three possible three-spheres in $X$. They all have a $U(1)$ isometry
whose orbit stabilizes asymptotically and vanishes in the interior over
a three-sphere.
Hence the M theory background
${\bf R}^4 \times X$ can be reduced to type IIA where it corresponds
to the full metric of one or a group of $N$ D6 branes wrapped on
the blown up three-sphere inside the deformed conifold.
The second class describes a completely new class of solutions
which, similarly to the first class, has a $U(1)$ isometry whose
orbit stabilizes at large radii. But in the interior the behaviour
is quite different because the orbit does not go to zero but is
of finite size everywhere. In type IIA these solutions correspond
to backgrounds without D6 branes but with RR two-form flux over
the blown up two-sphere inside the resolved conifold
$\mathcal{O}(-1) \oplus \mathcal{O}(-1) \to {\bf P}^1$.
The third class of solutions is a new branch of solutions of the
$G_2$ conditions of the first class and were found in \cite{cvnew}.
Locally the equations are the same but the manifold is different
globally. In this case one finds a metric on the ${\bf R}^2$ bundle
over $T^{11}$ which asymptotically looks like a $U(1)$ bundle over
a ${\bf Z}_2$ orbifold of the deformed conifold. The interesting
feature of this metric is the existence of a $U(1)$ isometry
with everywhere finite size orbit \cite{cvnew}, like in the second
class of solutions we found. Hence, also this corresponds to a type
IIA vacuum solution with pure RR two-form flux. In this case the
flux is over the four-cycle ${\bf P}^1 \times {\bf P}^1$ inside
the line-bundle over ${\bf P}^1 \times {\bf P}^1$ \cite{pagepope}.

What remains as a challenge is to find analytic solutions
for the last two classes of metrics which reduce to
the pure {\it flux solutions} in type IIA.
So far we were only able to find numerical evidence for the
existence of these metric.
Combined with the knowledge of asymptotic solutions
and solutions in the interior to high orders this evidence
is very strong but it would be nice to get better analytic
control. In this respect we hope that our approach will
turn out to be useful, too,
since it reduces the problem to a minimum of unknown functions
and a single second-order differential equation.
Another application would be to study metrics on
other manifolds known in the literature. For example
one could try to generalize the metrics on the ${\bf R}^3$ bundles
over ${\bf S}^4$ and ${\bf P}^2$ found by \cite{bs,gpp}.
These would correspond to the $M$-theory lift of the full
solutions\footnote{Note that in this context the solutions in \cite{bs,gpp}
correspond to the near horizon limit.} of localized intersecting D6 branes.
Furthermore, in a recent paper \cite{achwit} new singular conical
$G_2$ metrics were conjectured to exist on spaces which are cones
over weighted projective spaces
${\bf WP}_{[n,n,m,m]}$. Naturally it would be interesting to
find the explicit $G_2$ structure on those spaces.

\section*{Acknowledgments}

I would like to thank J.~Gomis, K.~Sfetsos and E.~Witten for discussions.
This work was supported in part by the DOE under grant No. DE-FG03-92ER40701.

While this paper was being written,
preprint \cite{xxx} appeared where the metric \eqn{fluxmetric} in
section 3.2 is found independently.

\newpage

\appendix

\section{General $SU(2) \times SU(2) \times U(1)$ symmetric ansatz}

In this appendix we present the general form of the
three-form ansatz and the corresponding metric and
second order equation for $G_2$ holonomy used in section 3.
The ansatz that we consider has $SU(2)_L \times \widetilde{SU(2)}_L
\times U(1)_R^\mathrm{diag}$ symmetry
\ba\label{app01}
\Phi & = & r_0^3 \left( p \, \sigma_1 \wedge \sigma_2 \wedge \sigma_3 +
q \, \Sigma_1 \wedge \Sigma_2 \wedge \Sigma_3 \right) \nonumber \\
 & & \quad + \, d \left( a(r) (\sigma_1 \wedge \Sigma_1 +
\sigma_2 \wedge \Sigma_2) + b(r) \sigma_3 \wedge \Sigma_3 \right) ~,
\ea
where $U(1)_R^\mathrm{diag} \sim SO(2)$ acts diagonally from the right on the
left-invariant one-forms $\sigma_{1,2}$ and $\Sigma_{1,2}$.
The metric resulting from \eqn{g2metric} and \eqn{app01} is
\ba\label{app02}
ds^2 & = & \bigg[ a (b - p r_0^3) a' \, \Big( (\sigma_1)^2 +
  (\sigma_2)^2 \Big) \nonumber \\
 & & -(b^2 + p q r_0^6) a' \, \Big( \sigma_1 \Sigma_1 + \sigma_2 \Sigma_2
\Big) \nonumber \\
 & & + a (b + q r_0^3) a' \, \Big( (\Sigma_1)^2 +  (\Sigma_2)^2
\Big) \nonumber \\
 & & + (a^2 - p r_0^3 b) b' \, \Big( \sigma_3 \Big)^2 \nonumber \\
 & & + (b^2 - 2 a^2 - p q r_0^6) b' \, \Big( \sigma_3 \Sigma_3 \Big)
\nonumber \\
 & & + (a^2 + q r_0^3 b) b' \, \Big( \Sigma_3 \Big)^2
 + (a')^2 b' dr^2 \bigg]/\Omega ~,
\ea
with
\be\label{app03}
\Omega = \frac{(a')^{2/3} (b')^{1/3}}{2^{2/3}}
\Big[ 4 a^2 (b - p r_0^3)(b + q r_0^3) -
(p q r_0^6 + b^2)^2 \Big]^{1/3} ~.
\ee
The functions $a$ and $b$ obey a second order differential
equation
\ba\label{app04}
&& 4 a' b' \Big[a a' (b-p r_0^3) (b +q r_0^3) + \nonumber \\
&& b' (a^2((p-q) r_0^3 - 2 b) + p q r_0^6 b + b^3) \Big] + \nonumber \\
&& (a' b'' - a'' b') \Big[ 4 a^2 (p r_0^3 - b)(q r_0^3 + b) +
(p q r_0^6 + b^2)^2 \Big] = 0 ~.
\ea

\section{A $SU(2) \times SU(2)$ symmetric ansatz}

In this appendix we present a three-form ansatz with
$SU(2)_L \times \widetilde{SU(2)}_L$ symmetry
and the corresponding metric and
second order equation for $G_2$ holonomy.
Note that contrary to \eqn{app01} this is not
the most general three-form ansatz consistent
with the symmetries. It is merely a natural generalization of
the ansatz \eqn{app01} by introducing one additional
function which breaks the $U(1)_R^\mathrm{diag}$ symmetry.
The three-form ansatz is
\ba\label{app11}
\Phi & = & r_0^3 \left( p \, \sigma_1 \wedge \sigma_2 \wedge \sigma_3 +
q \, \Sigma_1 \wedge \Sigma_2 \wedge \Sigma_3 \right) \nonumber \\
 & & \quad + \, d \left(
a(r) \sigma_1 \wedge \Sigma_1 +
b(r) \sigma_2 \wedge \Sigma_2 +
c(r) \sigma_3 \wedge \Sigma_3 \right) ~.
\ea
The metric resulting from \eqn{g2metric} and \eqn{app11} is
\ba\label{app12}
ds^2 & = &
 \bigg[ (b c - p r_0^3 a) a' \, \Big( \sigma_1 \Big)^2 +
        (b c + q r_0^3 a) a' \, \Big( \Sigma_1 \Big)^2
\nonumber \\
 & & + (a^2 - b^2 - c^2 - p q r_0^6) a' \, \Big( \sigma_1 \Sigma_1 \Big)
\nonumber \\
 & & + (a c - p r_0^3 b) b' \, \Big( \sigma_2 \Big)^2 +
       (a c + q r_0^3 b) b' \, \Big( \Sigma_2 \Big)^2
\nonumber \\
 & & + (b^2 - a^2 - c^2 - p q r_0^6) b' \, \Big( \sigma_2 \Sigma_2 \Big)
\nonumber \\
 & & + (a b - p r_0^3 c) c' \, \Big( \sigma_3 \Big)^2 +
       (a b + q r_0^3 c) c' \, \Big( \Sigma_3 \Big)^2
\nonumber \\
 & & + (c^2 - a^2 - b^2 - p q r_0^6) c' \, \Big( \sigma_3 \Sigma_3 \Big)
     + a' b' c' dr^2 \bigg]/\Omega ~,
\ea
with
\ba\label{app13}
\Omega & = & -\frac{(a' b' c')^{1/3}}{2^{2/3}}
\Big[ a^4 + b^4 + c^4 - 2 (a^2 b^2 + a^2 c^2 + b^2 c^2) \nonumber \\
 & & + 4 (p-q) r_0^3 a b c +
2 p q r_0^6 (a^2 + b^2 + c^2) + p^2 q^2 r_0^{12}
\Big]^{1/3} ~.
\ea
The functions $a$, $b$ and $c$ obey a set of two second order
differential equations
\ba\label{app14}
&& a' (b' (4 a' (-a^3 a' + b (- (p-q) r_0^3 c a' + 2 (p q r_0^6 + \nonumber \\
&&  b^2 - c^2) b') - c (p q r_0^6 - b^2 + c^2) c' + \nonumber \\
&& a^2 (-2 b b' + c c') + a ((-p q r_0^6 + b^2 + c^2) a' + \nonumber \\
&& (p-q) r_0^3 (2 c b' - b c'))) - \nonumber \\
&& (a^4 + b^4 + 4 (p-q) r_0^3 a b c + 2 b^2 (p q r_0^6 -c^2) + \nonumber \\
&& 2 a^2 (p q r_0^6 - b^2 - c^2) + (p q r_0^6 + c^2)^2) a'') + \nonumber \\
&& 2 ( a^4 + b^4+4(p-q) r_0^3 a b c+2 b^2(p q r_0^6 - c^2)+ \nonumber \\
&& 2 a^2 (p q r_0^6 - b^2 -c^2)+(p q r_0^6+c^2)^2 a' b'')- \nonumber \\
&& (a^4 + b^4 + 4(p-q) r_0^3 a b c + 2 b^2(p q r_0^6-c^2)+ \nonumber \\
&& 2 a^2(p q r_0^6 -b^2-c^2)+(p q r_0^6 + c^2)^2) a' b' c'') \nonumber \\
&& = 0 ~,
\ea
and
\ba\label{app15}
&& b'(c'(2 b'(2 a'(2 a^3 a'+b(2(p-q) r_0^3 c a'-(p q r_0^6+ \nonumber \\
&& b^2 -c^2)b')-c(p q r_0^6 -b^2+c^2)c'+ \nonumber \\
&& a^2 (b b' + c c') + a (2(p q r_0^6 - b^2 -c^2)a' - \nonumber \\
&& (p-q) r_0^3 (c b' + b c'))) + \nonumber \\
&& (a^4 +b^4+4(p-q) r_0^3 a b c+2 b^2(p q r_0^6-c^2)+ \nonumber \\
&& 2 a^2 (p q r_0^6-b^2 -c^2)+(p q r_0^6 +c^2)^2)a'')- \nonumber \\
&& (a^4+b^4+4(p-q) r_0^3 a b c + 2 b^2(p q r_0^6-c^2)+ \nonumber \\
&& 2 a^2(p q r_0^6-b^2-c^2)+(p q r_0^6+c^2)^2)a' b'')- \nonumber \\
&& (a^4+b^4+4(p-q) r_0^3 a b c +2 b^2 (p q r_0^6-c^2)+ \nonumber \\
&& 2 a^2(p q r_0^6 - b^2-c^2)+(p q r_0^6+c^2)^2)a' b' c'')   \nonumber \\
&& = 0 ~.
\ea
Again one of the functions $a$ , $b$ , $c$ or a combination of them
can be eliminated using reparametrization invariance in $r$.

\end{document}